\documentstyle[prb,aps,epsf]{revtex}
\begin{document}
\title{Phase diagram of an extended Kondo lattice model for 
manganites: the Schwinger-boson mean-field approach}
\author{R. Y. Gu, Z. D. Wang, Shun-Qing Shen}
\address{Department of Physics, The University of Hong Kong, Pokfulam Road, Hong Kong}
\author{D. Y. Xing}
\address{National Laboratory of Solid State Microstructures and
Department of Physics, Nanjing University, Nanjing 210093, China}
\date{\today}

\twocolumn[\hsize\textwidth\columnwidth\hsize\csname@twocolumnfalse\endcsname
\maketitle

\begin{abstract}
We investigate the phase diagram of an extended Kondo lattice model for
doped manganese oxides in the presence of strong but finite Hund's
coupling and on-site Coulomb interaction. By means of the Schwinger-boson
mean-field approach, it is found that, besides magnetic ordering, there
will be non-uniform charge distributions, such as charge ordering and
phase separation, if the interaction between electrons prevails over the
hybridization. Which of the charge ordering and phase separation appears
is determined by a competition  
between effective repulsive and attractive
interactions due to virtual processes of electron hopping. Calculated
results show that strong electron correlations caused by the
on-site Coulomb interaction as well as the
finite Hund's coupling play an
important role in the magnetic ordering and charge distribution at
low temperatures.
\end{abstract}


]

\section{Introduction}

The doped manganese oxides with perovskite structure, R$_{1-x}$A$_x$MnO$_3$
(R=La, Pr, Nd; and A=Sr, Ca, Ba, Pb), have recently attracted much
attention due to the observation of colossal magnetoresistance (CMR).
\cite{helmolt93,jin94,tokura94,fontcuberta96} Theories based on the double
exchange (DE) model, \cite{zener51,anderson55,gennes60} in which the
exchange of electrons between neighboring manganese ions with a strong
Hund's coupling drives spins of on-site electrons to align parallelly,
have been developed for a long time and can qualitatively elucidate the
relation of transport and magnetism in the doping range of $0.2<x<0.45$.
However, recent systematic experimental studies revealed rich phase diagrams,
which are difficult to be understood by the DE model alone. For instance,
the system is actually insulating in the half-doping case $(x=0.5)$
at low temperature; but a metallic ferromagnetic (FM) state
would be expected according to the DE model, for the DE hopping reaches
its maximum and the effective FM
interaction becomes strongest at this doping. Furthermore,
for $x= 0.5$, a charge ordering characterized by an alternating
Mn$^{3+}$ and Mn$^{4+}$-ion arrangement in the real space was observed
to be superimposed on the antiferromagnetic (AF) ordering.
\cite{goodenough55,kuwahara95} This charge ordering is sensitive to an
applied magnetic field, and even melts under a moderate magnetic field. 
In the meantime, the resistance may decrease by several
orders in magnitude, \cite{kuwahara95,tomioka}  implying that there is a
close relation between the charge ordering and the AF spin
background.

Many efforts have been made to 
understand the phase diagram of 
a doped manganese oxide  based on
various models.\cite{inoue95,shiina,ishihara,brink,arovas98} To
explore the origin of the unique magnetotransport, many further
theories on the basis of DE model have been proposed, such as Jahn-Teller
displacement and electron-phonon interactions,\cite{millis,alexandrov}
spin-polaron formation,\cite{palstra} Anderson localization with diagonal
and off-diagonal disorder, \cite{sheng,zhong} and the phase separation
scenario.\cite{yunoki98,shen98}
Importance of various interactions on the physical
properties of the manganites is currently one of the lively-debating
subjects. A comprehensive understanding of the magnetic and charge
ordering states as well as their relations to the transport properties are
highly desirable.

In this paper, we investigate effects induced by strong but
finite Hund's coupling $J_H$ and on-site Coulomb interaction $U$.
In the manganites, three $t_{2g}$ electrons are almost localized and form
an $S=3/2$ spin state according to the Hund's rule, while electrons in
$e_g$ orbit evolve a conduction band. In the conventional DE model $J_H$
is usually regarded to be infinite and $U$ is neglected, so that there
exist only single-occupied state with spin $S+1/2$ and empty state with
spin $S$. In this limit, since neither $S-1/2$ state nor double occupancy
of $e_g$ electrons on the same site is allowed, the effects induced by two
kinds of virtual processes, that an electron hops to a empty site with its
spin antiparallel to the core spin on the site and that an electron
hops to a singly-occupied site, are completely eliminated. In this work we
will show that the effects due to the two kinds of virtual processes are
important to account for the observed phase diagrams. On one hand,
the two kinds of virtual processes can produce an AF superexchange coupling
between the neighboring localized spins, with coupling strength being
$t^2/(2J_H)$ and $t^2/(J_H+U)$, respectively, where $t$ is the hopping
integral of the $e_g$ electron. The AF coupling induced by the virtual
processes is usually much stronger than the direct AF superexchange
coupling between neighboring localized spins. On the other hand,
the virtual process of producing a single-occupied state with spin
$S-1/2$ can lead to a repulsive interaction between conduction electrons,
while the virtual process of producing a double occupancy can result
in an attractive one, as will be shown in Sec. IIA.
The two types of interactions between electrons can produce non-uniform
charge distributions,  such as charge ordering and electronic phase
separation, provided that the interactions prevail over the
hybridization effects of electrons. Which of charge ordering and phase
separation appears is determined by a competition at different doping
between two types of
interactions, for the repulsive interaction is favorable to the charge
ordering while the attractive interaction may cause the phase separation.

In Sec. II, starting from an extended Kondo lattice model, we have an
effective projected Hamiltonian in the case of strong but finite $J_H$
and $U$. The repulsive and attractive interactions, which are associated
with magnetic ordering and non-uniform charge distributions, are obtained.
A Schwinger-boson mean-field theory is developed to establish the phase
diagram at low temperatures. In Sec. III we focus attention on the case
of $x=1/2$. The possibility that the FM, AF or canted ferromagnetic (CF)
order appears, as well as that the Wigner lattice or phase separation
appears, is discussed. In Sec. IV numerical results for phase digram
are presented. In the large $J_H$ case, the hybridization effect is
dominant and there exists metallic ferromagnetism, which accords with
the DE model. As the repulsive interaction due to finite $J_H$ effects
is relatively strong, the charge ordering will be formed at half doping in
the AF background. On the other hand, the phase separation may arise when
the attractive interaction due to finite $U$ becomes dominant.

\section{General formalism}

\subsection{Effective projected Hamiltonian}

The electronic model Hamiltonian for doped manganese oxides we 
considered presently is given by
\begin{eqnarray}
H &=&-t\sum_{(ij),\sigma }c_{i\sigma }^{\dag }c_{j\sigma
}+U\sum_{i}n_{i\uparrow }n_{i\downarrow }  \nonumber \\
&&-J_{H}\sum_{i\sigma \sigma ^{\prime }}c_{i\sigma }^{\dag }{\bf S}_{i}\cdot 
{\mbox {\boldmath $\sigma$}}_{\sigma \sigma ^{\prime }}c_{i\sigma
^{\prime }}+J_{AF}\sum_{(ij)}{\bf S}_{i}\cdot {\bf S}_{j}\ .
\end{eqnarray}
Here $c_{i\sigma }$ ($c_{i\sigma }^{\dag}$) is the annihilation (creation)
operator for conduction electrons at site $i$ with spin $\sigma$,
$n_{i\sigma}=c_{i\sigma }^{\dag }c_{i\sigma}$ is the particle number
operator, ${\bf S}_{i}$ is the total spin operator of the localized
electrons at site $i$ with $S=3/2$, and {\boldmath $\sigma$}
is the Pauli matrix. In the manganites the Hund's coupling constant
$J_{H}$ and the on-site Coulomb interaction $U$ are much greater than the 
hopping integral $t$ as well as the direct AF superexchange coupling
$J_{AF}$. On a given site, an itinerant electron constrained by the strong
Hund's coupling has its spin parallel to the core spin, forming a spin
$S+1/2$ state.  The singly-occupied state for the itinerant electron with
spin antiparallel to the core spin and the doubly-occupied state are almost
forbidden, making it appropriate to utilize the projective perturbation
technique to investigate Hamiltonian (1). The effects of finite $J_{H}$ and
$U$ can be regarded as a perturbation correction to the large $J_{H}$ and
$U$ limit where there are only the empty and single occupancy with
$S+1/2$ state. Up to the second-order perturbation, an effective
Hamiltonian for Eq.\ (1) can be derived as\cite{shen97}  
\begin{eqnarray}
H_{eff} &=&-t\sum_{(ij),\sigma }\overline{c}_{i\sigma }^{\dag }\overline{c}%
_{j\sigma }+J_{AF}\sum_{(ij)}\overline{{\bf S}}_{i}\cdot \overline{{\bf S}}%
_{j}  \nonumber \\
&&+J_{1}\sum_{(ij)}({\bf S}_{i}\cdot \tilde{{\bf S}}_{j}-S\tilde{S}%
)P_{ih}P_{js}^{\dag }  \nonumber \\
&&+J_{2}\sum_{(ij)}(\tilde{{\bf S}}_{i}\cdot \tilde{{\bf S}}_{j}-\tilde{S}%
^{2})P_{is}^{\dag }P_{js}^{\dag }\ .
\end{eqnarray}
Here ${\bf \tilde{S}}_{i}$ is the spin operator with $\tilde{S}=S+1/2$. $%
P_{ih}$ and $P_{is}^{\dag }$ are the projection operators for empty state
with spin $\bf S_i$ and single occupancy with spin $\bf \tilde{S}_i$,
respectively. $\overline{{\bf S}}_{i}={\bf S}_{i}P_{ih}+S\tilde{{\bf S}}_{i}
P_{is}^{\dag }/\tilde{S}$. $\overline{c%
}_{i\sigma}=P_{i}c_{i\sigma}P_{i}$ and
 $\overline{c}_{i\sigma}^{\dag }=P_{i}c_{i\sigma}^{\dag }P_{i}$
are projected electron operators, where $P_{i}=P_{ih}+P_{is}^{\dag }$, 
projects onto the space of non-doubly-occupied site.
The last two terms in Eq.\ (2) are the second-order
perturbation corrections where $J_{1}=t^{2}/(2J_{H}\tilde{S}^{3})$ and
$J_{2}=t^{2}/[(J_{H}S+U)\tilde{S}^{2}]$, respectively, stemming from
different virtual processes (a) and (b). In virtual process (a), an electron
first hops from a site to one of the nearest-neighbor empty sites to form a
spin $S-1/2$ state and then hops backward; while in virtual process (b),  
an electron first hops to a single-occupied site, where there has been
an electron with opposite spin, and then hops backward. Owing to $J_1>0$
and $J_2>0$, both virtual processes favor the AF arrangement of the core
spins and enhance greatly the AF coupling between the neighboring spins.
Whether non-uniform charge distributions can
appear is dertermined by a competition between electronic hybridization
and interactions of electrons with one another. The hybridization, the
overlap of electron wavefunctions centred on different sites, is a
quantum-mechanical effect that allows electrons to hop from one atom to
another, thus tending to spread the electronic density uniformly through
the system. In contrast, the interactions of electrons in the system tend
to promote non-uniform charge distributions.\cite{nature} The addition
of the last two terms in Eq.\ (2), arising from the finite $J_H$ and $U$
effects, enhances greatly the interaction side in the competition and
thus favors
the occurance of the non-uniform charge distributions. On the other hand,
since the values of 
$J_1({\bf S}_{i}\cdot \tilde{{\bf S}}_{j}-S\tilde{S})$ and
$J_2(\tilde{{\bf S}}_{i}\cdot \tilde{{\bf S}}_{j}-\tilde{S}^{2})$ are
always non-positive,
it is clear that the two terms
represent the repulsive and attractive interactions between conduction
electrons on neighboring sites, respectively. The
competition between them will lead to different non-uniform charge
distributions, the charge-ordered state and the phase separation.

\subsection{Schwinger boson representation}

A projected electron operator may be regarded as a combination of
the operator $f$ for a spinless charge fermion and that for a neutral
boson with spin $\tilde{S}=S+1/2$, i.e.,
$\overline{c}_{i\sigma }=f_{i}b_{i\sigma }/\sqrt{2\tilde{S}}$, where
$b_{i\sigma }$ is the Schwinger boson operator. \cite{Auerbach} 
In the Schwinger boson representation, the spin
${\bf S}_{i}^{\prime }$ $({\bf S^{\prime }}_{i}=%
{\bf S}_{i}$ or ${\bf \tilde{S}}_{i})$ can be expressed
as ${\bf S}_{i}^{\prime }=\frac{1}{2}\sum_{\sigma \sigma
^{\prime }}b_{i\sigma }^{\dag }{\mbox {\boldmath $\sigma$}}%
_{\sigma \sigma ^{\prime }}b_{i\sigma ^{\prime }}$, with constraint $%
\sum_{\sigma }b_{i\sigma }^{\dag }b_{i\sigma }=2S^{\prime }$ ($S^{\prime }=S$
or $\tilde{S}$). This constraint can also be written as $\sum_{\sigma
}b_{i\sigma }^{\dag }b_{i\sigma }=2S+n_{i}$, where $n_{i}=f_{i}^{\dag }f_{i}$
is the particle number operator of fermions. The projection operators can
be replaced by $P_{ih}=1-n_{i}$ and $P_{is}^{\dag }=n_{i}$.
Then, by using the identity ${\bf S}_{i}^{\prime
}\cdot {\bf S}_{j}^{\prime }=-\frac{1}{2}A_{ij}^{\dag }A_{ij}+S_{i}^{\prime
}S_{j}^{\prime }$ with $A_{ij}=\frac{1}{2}(b_{i\uparrow }b_{j\downarrow
}-b_{i\downarrow }b_{j\uparrow })$, the effective Hamiltonian is reduced to 
\begin{eqnarray}
H &=&J_{AF}S^{2}-\tilde{t}\sum_{ij}f_{i}^{\dag }f_{j}F_{ij} \nonumber \\
&\ &-\frac{1}{2}A_{ij}^{\dag }A_{ij}[J_{AF}+\tilde{J_{1}}(1-n_{i})n_{j}+\tilde{%
J_{2}}n_{i}n_{j}]\ ,
\end{eqnarray}
where $F_{ij}=\sum_{\sigma }b_{i\sigma }^{\dag }b_{j\sigma }$, $\tilde{t}%
=t/(2\tilde{S})$, $\tilde{J_{1}}=J_{1}-J_{AF}/\tilde{S}$ and $\tilde{J_{2}}%
=J_{2}-J_{AF}/\tilde{S}+J_{AF}/4\tilde{S}^{2}$. The difference between $%
\tilde{J}_{n}$ and $J_{n}$ ($n=$1 and 2) is caused by the finite
$S$ effect, and would disappear in the large spin limit.

\subsection{Mean-field approximation}

We now make a Hartree-Fock mean-field approximation by 
decoupling various terms in Eq.(3). As an example, we have
\begin{eqnarray}
A_{ij}^{\dag }A_{ij}n_{i}n_{j} &\rightarrow &A_{ij}^{\dag
}<A_{ij}><n_{i}><n_{j}>  \nonumber \\
+ &<&A_{ij}^{\dag }>A_{ij}<n_{i}><n_{j}>  \nonumber \\
+ &<&A_{ij}^{\dag }><A_{ij}>n_{i}<n_{j}>  \nonumber \\
+ &<&A_{ij}^{\dag }><A_{ij}><n_{i}>n_{j}  \nonumber \\
-3 &<&A_{ij}^{\dag }><A_{ij}><n_{i}><n_{j}>\ . \nonumber
\end{eqnarray}
For the fermions,
we have $<n_{i}>=1-x$ in a uniform density state. In the presence of the
charge ordering at $x=1/2$, the lattice can be divided into two sublattices
$X=A$ and $B$, and $<n_i>=N_{X}=1-x+\alpha\delta$ where
$\alpha=1$ or $-1$ for $i\in A$ or B,
and $\delta$ $(0\leq \delta \leq 1-x)$ is the charge ordering parameter.
The uniform state can be regarded as a special case of $\delta=0$.
As in Ref.\cite{sarker,arovas98}, the FM and AF order parameters 
can be respectively written as
\begin{eqnarray*}
<F_{ij}> &=&F  \label{def-2} \ , \\
<A_{ij}> &=&(-1)^{r_{i}}A\ ,  
\end{eqnarray*}
where 
$r_{i}$ depends on the position of site $i$, being an odd number in one
sublattice and an even number in the other sublattice. The average value of
$<f_{i}^{\dag }f_{j}>$ is assumed to be a constant $K$.
Under the Hartree-Fock mean-field approximation, Hamiltonian (3) is
reduced to
\begin{equation}
H_{MF}=E_{0}+H_{Bose}+H_{Fermi}
\end{equation}
with 
\begin{eqnarray*}
E_{0} &=&NZ\tilde{t}KF+NZJ_{AF}S^{2}-N\Lambda 2S  \\
&&+\frac{1}{2}%
NZA^{2}\{J_{AF}+2(1-x)\tilde{J}_{1} \nonumber \\
&&+3[( 1-x) ^{2}-\delta ^{2}](\tilde{J}_{2}-\tilde{J}%
_{1})\}\, \\
H_{Bose} &=&\sum_{(ij)}\frac{-\tilde{J}}{2}(-1)^{r_{i}}A(A_{ij}^{\dag
}+A_{ij}) \\
&&+\Lambda \sum_{i\sigma}b_{i\sigma }^{\dag }b_{i\sigma }-\tilde{t}%
K\sum_{(ij)\sigma}b_{i\sigma }^{\dag }b_{j\sigma } \ , \\
H_{Fermi} &=&-\tilde{t}F\sum_{(ij)}f_{i}^{\dag }f_{j}-(\Lambda +\mu
)\sum_{i}n_{i} \\
&&-\frac{1}{2}A^{2}Z\sum_{i}[2(\tilde{J}_{2}-\tilde{J}_{1})%
N_{\bar{X}}+\tilde{J}_{1}]n_{i}\ .
\end{eqnarray*}
Here $\tilde{J}=J_{AF}+(\tilde{J}_{2}-\tilde{J}_{1})[(1-x)^{2}-\delta ^{2}]
+\tilde{J}_{1}(1-x)$ is the effective AF coupling constant, $N$ is the
number of lattice sites, $Z$ is the coordinate number of the lattice,
and $N_{\bar{X}}=N_{B}$ ($N_A$) for $i\in A$ ($B$).
$\Lambda=<\Lambda_i>$ with $\Lambda_i$ being the Lagrange multiplier
introduced at site $i$ to enforce the local
constraint $\sum_{\sigma }b_{i\sigma }^{\dag }b_{i\sigma }=2S+n_{i}$,\cite{Auerbach} 
and $\mu $ is the chemical potential of fermions which is determined by
the total conduction electron number.  After
diagonalizing $H_{Bose}$ and $H_{Fermi}$, we have 
\[
H_{Bose}=\sum_{{\bf k}\sigma }E_{{\bf k}}\beta _{{\bf k}\sigma }^{\dag
}\beta _{{\bf k}\sigma }+\sum_{{\bf k}}(E_{{\bf k}}-\Lambda )\ ,
\]
and
\[
H_{Fermi}=\sum_{{\bf k}}(\varepsilon _{{\bf k}}-\tilde{\mu}%
)\psi _{{\bf k}}^{\dag }\psi _{{\bf k}}\ ,
\]
with boson and fermion dispersions as
\[
E_{{\bf k}}=\sqrt{\Lambda ^{2}-(ZA\tilde{J}\gamma _{{\bf k}})^{2}}-Z\tilde{t}%
K\gamma _{{\bf k}}\ ,
\]
and
\begin{eqnarray*}
\varepsilon _{{\bf k}}=-sgn(\gamma_{{\bf k}})Z\sqrt{%
A^{4}(\tilde{J}_{2}-\tilde{J}_{1})^{2}\delta ^{2}+\tilde{t}^{2}F^{2}\gamma
_{{\bf k}}^{2}}\ , 
\end{eqnarray*}
respectively. Here
$\gamma _{{\bf k}}=Z^{-1}\sum_{{\mbox {\boldmath $\eta$ }}}e^{i{\bf k}
\cdot {\mbox {\boldmath $\eta$ }}}$
with {\boldmath $\eta$} the vector of the nearest neighbors of each site, and
the shifted chemical potential is given by
\begin{eqnarray*}
\tilde{\mu}=\mu +\Lambda +\frac{1}{2}ZA^{2}[
\tilde{J}_{1}+2(\tilde{J}_{2}-\tilde{J}_{1})(1-x)]\ .
\end{eqnarray*}
From the fermion spectrum, it is found that the energy band of fermions
is divided into two separated branches, between which there is an energy
gap provided $\delta\neq 0$ and $\tilde{J}_2\neq \tilde{J}_1$.
Thus, in the half-doping case ($x=1/2$) at zero temperature, the lower
branch ($\gamma_{{\bf k}}>0$) is fully occupied by fermions while the
upper one ($\gamma_{{\bf k}}<0$) is empty.

\section{Charge ordering and phase separation}

Now we come to discuss the magnetic ordering and non-uniform
charge distributions at $x=0.5$. First let us
write down the mean-field equations. When
the magnetic ordering arises, the Schwinger bosons should
condensate to the lowest energy state. From $E_{{\bf k}=0}=0$, we have 
\[
\Lambda =Z\sqrt{\tilde{t}^{2}K^{2}+A^{2}\tilde{J}^{2}}\ .
\]
Using the spectra of quasi-fermions and bosons, we obtain a set of
self-consistent equations at zero temperature: 
\begin{mathletters}
\begin{eqnarray}
K &=&\frac{1}{2N}\sum_{{\bf k}}\frac{\tilde{t}F\gamma _{{\bf k}}^{2}}{\sqrt{(%
\tilde{J}_{1}-\tilde{J}_{2})^{2}A^{4}\delta ^{2}+\tilde{t}^{2}F^{2}\gamma _{%
{\bf k}}^{2}}}\ ,   \\
F &=&\frac{\tilde{t}K(2S+2-x)}{\sqrt{\tilde{t}^{2}K^{2}+A^{2}\tilde{J}%
^{2}}} \nonumber \\
&&-\frac{1}{N}\sum_{{\bf k}}\frac{\tilde{t}K}{\sqrt{\tilde{t}^{2}K^{2}+A^{2}%
\tilde{J}^{2}(1-\gamma _{{\bf k}}^{2})}}\ ,   \\
A &=&\frac{A\tilde{J}(2S+2-x)}{\sqrt{\tilde{t}^{2}K^{2}+A^{2}\tilde{J}%
^{2}}}  \nonumber \\
&&-\frac{1}{N}\sum_{{\bf k}}\frac{A\tilde{J}(1-\gamma _{{\bf k}}^{2})}{\sqrt{\tilde{t%
}^{2}K^{2}+A^{2}\tilde{J}^{2}(1-\gamma _{{\bf k}}^{2})}}\ ,  
\\
\label{self-4}
\delta  &=&\frac{1}{2N}\sum_{{\bf k}}\frac{(\tilde{J}_{1}-\tilde{J}%
_{2})A^{2}\delta }{\sqrt{(\tilde{J}_{1}-\tilde{J}_{2})^{2}A^{4}\delta ^{2}+%
\tilde{t}^{2}F^{2}\gamma _{{\bf k}}^{2}}}\ .  
\end{eqnarray}
\end{mathletters}
From Eq.\ (\ref{self-4}),  it follows that the charge ordering may appear $%
(\delta \neq 0)$ only when $\tilde{J}_{1}>\tilde{J}_{2}$, otherwise
there exists only a trivial solution of $\delta =0$. Physically, 
the effective interaction between electrons is repulsive for $\tilde{J}%
_{1}-\tilde{J}_{2}>0$, which is responsible for the formation of 
the charge ordering. There are two types of possible ordering states:
the pure magnetic ordering phase without charge order and the charge
ordering phase in which the charge and magnetic orders coexist.
The former includes FM, AF, and CF
phases; while the latter is the Wigner lattice ($\delta =1/2$, $K=F=0$,
and $A=A_{max}$) or a combination of the charge ordering
($0<\delta <1/2 $) and magnetic ordering ($A<A_{max}$, $K$ and $F\neq 0$).
Since the system under consideration is a three-dimensional (3D) simple-cubic
lattice, for convenience, we introduce two parameters
$\eta_{1}=\tilde{t}^{2}F^{2}/[(\tilde{J}_{1}-\tilde{J}_{2})^{2}A^{4}\delta
^{2}]$ $(0\leq \eta _{1}\leq \infty )$ and $\eta _{2}=-A^{2}\tilde{J}%
^{2}/(\tilde{t}^{2}K^{2}+A^{2}\tilde{J}^{2})$ $(-1\leq \eta _{2}\leq 0)$ and
define the following two 3D integrals: 
\begin{eqnarray}
C_{i} &=&\frac{1}{N}\sum_{{\bf k}}\frac{1}{\sqrt{1+\eta _{i}\gamma _{{\bf k}%
}^{2}}}\ ,  \nonumber \\
D_{i} &=&\frac{1}{N}\sum_{{\bf k}}\frac{\gamma _{{\bf k}}^{2}}{\sqrt{1+\eta
_{i}\gamma _{{\bf k}}^{2}}}\ .  \nonumber
\end{eqnarray}
with $i=1, 2$. 
Then the set of mean-field equations (5a)-(5d) can be rewritten as 
\begin{mathletters}
\begin{eqnarray}
K &=&\frac{1}{2}\sqrt{\eta _{1}}D_{1}\ ,   \\
F &=&\sqrt{1+\eta _{2}}(2S+2-x-C_{2})\ ,   \\
A &=&\sqrt{-\eta _{2}}(2S+2-x-C_{2}+D_{2})\ ,  \\
\label{delta}
\delta  &=&\frac{1}{2}C_{1}  \ .
\end{eqnarray}
\end{mathletters}
For the Wigner lattice ($\delta =1/2$), from Eq. (\ref{delta}), it follows
$C_1=1$ and $\eta _{1}=0$, so that $K=0$, $\eta _{2}=-1$, and
$F=0$. In this case, $A$ reaches its
maximum $A_{max}\approx 2S+0.597$, indicating a fully AF insulator
in which all the electrons are localiezed in one sublattice.
Another limiting case is that there is no charge ordering ($\delta =0$),
where there are two sets of solutions: (i) $K=F=0$ and $A=A_{max}$, which
is an AF state; and (ii) $\eta _{1}=+\infty$, $K\neq 0$, and $F\neq 0$,
which is a CF state for $0<A<A_{max}$ or a FM state for $A=0$.
There is the lowest energy in the ground state. This ground state energy
per site in the AF state [case (i)] is 
\begin{equation}
E_{g}=-\tilde{J}A_{max}^{2}/2\ ,  \label{eaf}
\end{equation}
and that for case (ii) is given by
\begin{equation}
E_{g}=-\frac{(2S+2-x-C_{2}-\eta _{2}D_{2})\tilde{t}}{6\sqrt{1+\eta _{2}}}-%
\frac{\tilde{t}^{2}\eta _{2}}{72\tilde{J}(1+\eta _{2})}\ ,  \label{ground}
\end{equation}
which is a function of  $\eta _{2}$. 
For a CF state, the magnitude of $\eta _{2}$ is
determined by minimizing $E_{g}$, 
i.e., $\partial E_{g}/\partial \eta _{2}=0$.
$\eta _{2}=0$ corresponding to a FM state.
It is found that  a phase transition occurs from a
CF state to a FM state when $\tilde{J}$ decreases to $%
\tilde{t}/(12S+9)$.

The uniform density phases discussed above may be unstable toward the
phase separation, which will occur if $\partial \mu/\partial x\leq 0$.  
In our model, when the virtual process (b) is dominant over process (a), i.e.,
$\tilde{J}_2>\tilde{J}_1$, the effective interaction between the $e_g$ 
electrons on neighboring sites will be attractive. If such an attractive
interaction is strong enough, the $e_g$ electrons will tend to accumulate
together to lower the total energy of the system. In this case, the
electronic density is no longer uniform and the phase separation occurs
between electron-rich and electron-poor regions. The chemical-potential
judgment of phase separation is $\partial \mu/\partial x \leq 0$, equivalent
to the energy judgment $\partial^2 E_{g}/\partial x^2 \leq 0$ due to
$\mu=\partial E_g/\partial x$. The phase separation will appear when
$\partial ^{2}E_{g}/\partial x^{2}\leq 0$, where both $\eta _{2}$
and $\tilde{J}$ in Eqs.\ (\ref{eaf}) and (\ref{ground}) are functions of
$x$.

\section{Phase diagrams and discussions}

Mean-field equations (6a)-(6d) have been numerically solved and the phase
diagrams
at $x=0.5$ are plotted in Figs.\ 1 and 2. To make the results comparable
with each other for different $S$, we have used reduced coupling constants
$j_{af}=J_{AF}S^{2}$ and $j_{h}=J_{H}\tilde{S}$. Fig. 1 is the phase
diagram of $j_{af}/t$ and $t/j_{h}$ in the large $U$ limit with
$S=3/2$ and $S\rightarrow \infty $. In this case, $J_2=0$, and
$t/j_h$ indicates the ratio of the repulsive interaction
due to virtue process (a) to the kinetic energy of the $e_g$ electron.
It is found that the system is in the FM state for small $j_{af}/t$ and
$t/j_{h}$, where either direct or indirect AF superexchange coupling is
weak compared with the FM coupling due to the DE mechanism. With increasing
$j_{af}/t$ and $t/j_{h}$, the AF superexchange coupling between the
neighboring localized spins is enhanced. As $\tilde{J}\geq \tilde{t}/(12S+9)$,
the CF state appears; meanwhile, a repulsive interaction between fermions
is increased due to the increase of $t/j_{h}$. At this stage, there is a CF
ordering but the charge ordering has not yet arisen, which is labeled
as the CF1 state in Figs.\ 1 and 2. When $t/j_{h}$ is increased beyond a
threshold so that the repulsive potential between fermions dominates over the
kinetic energy, the fermions begin to tend toward one sublattice and thus
charge ordering emerges, which is called as the CF2 state. Finally,
with further increase of $t/j_h$, $\delta$ increases gradually. At
$\delta =1/2$, all the $e_g$ electrons are confined in one sublattice in
the AF background, forming a Wigner lattice. It is found that the phase
diagram for $S=3/2$ is very similar to that for $S\rightarrow \infty$.
The difference between them can be seen in the large $j_{af}/t$ case,
for the effective repulsive interaction is proportional to the factor
$\tilde{J}_{1}-\tilde{J}_{2}$ or $t/j_{h}-j_{af}/(2tS^{2}\tilde{S})$.
For finite $S$, the increase of $j_{af}$ will decrease this factor
and so be unfavorable to the charge ordering. As a result, the quantum
fluctuation of finite $S$ leads to enlargements of the $CF1$ and $CF2$
regions.

For manganites, it is roughly estimated that $t\approx 0.15eV$, $j_{h}\approx
0.75eV$, $U\approx 10eV$, and $j_{af}\approx 8meV$, \cite{mishra}
so that $j_{af}/t\approx 0.05$ and $t/j_h\approx 0.2$. From Fig. 1,
it follows that the repulsive interaction from virtual process (a) alone
enables the $e_g$ electrons to form the charge ordering. This indicates that
this process plays an important role in determining the collective behavior
of electrons. In actual doped manganites, there may also be other effects
that favor charge ordered phase, which are not considered in the present
paper, such as the direct nearest-neighbor Coulomb interactions. \cite
{tomioka,mishra,lee} It is worth pointing out here whether the charge ordering
occurs depends not only the ratio of $t/j_h$, but also the magnetic
ordering of the system. In doped manganites, the amplitude for an electron
to hop from one site to another is determined by the relative orientation
between the core spins at the two sites, being greatest when the core spins
are parallel and least when they are antiparallel. As a result, a FM state
has the greatest hybridization and so tends to spread the electronic density
uniformly through the system; while an AF state has the least hybridization
and favors the charge ordering. This can account for the sensitivity
of charge ordering with respect to an applied magnetic field that
tends always to align the core spins. \cite{kuwahara95,tomioka} 
Besides, there is a difference between two types of repulsive
interactions mentioned above in the response to an applied magnetic
field: with a decrease in the AF correlation, the superexchange 
electron-electron interaction is reduced, while the direct Coulomb
interaction remains unchanged. Therefore, charge ordering caused by 
virtual process (a) is more easily affected by the magnetic field than
that caused by the direct Coulomb interaction.

\begin{figure}
\epsfxsize=8.5cm
\epsfbox{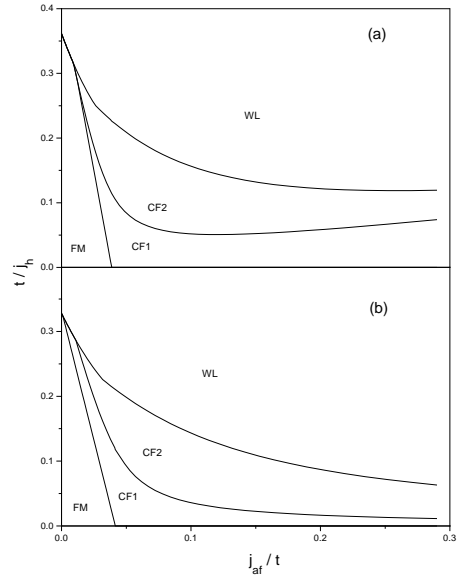}
\caption{Phase diagram of the extended Kondo lattice model with
       $U=\infty$ for (a) $S=1.5$ and (b) $S\rightarrow\infty$.
       FM, CF1, and CF2 denote regimes of metallic ferromagnet,
       canted ferromagnet without and with charge ordering, respectively.
       WL stands for the Wigner lattice.}
\end{figure}

In Fig. 2, we present the phase diagram of  $j_{h}/(j_{h}+U)$ and 
$t/j_{h}$.  In the finite $U$ case, $j_{h}/(j_{h}+U)$ describes the
relative strength of the effective interaction due to virtual process (b)
to that due to process (a). For small $t/j_h$, there is only a FM state,
where the Hund's coupling is very strong, both virtual processes (a) 
and (b) are suppressed so that the hybridization effect and the DE
ferromagnetism dominates over the system. With increasing $t/j_h$,
both $J_1$ and $J_2$ become large and so the AF superexchange coupling
due to virtue processes (a) and (b) is enhanced, leading to the CF1 state.
As $t/j_h$ is increased beyond a threshold, the charge ordering or phase
separation may occur, depending on the competition between effective
repulsive and attractive interactions. For $j_h<U$,
$\tilde{J}_{1}>\tilde{J}_{2}$, the net repulsive interaction favors the
charge ordering. The opposite case is $\tilde{J}_{1}<\tilde{J}_{2}$, 
where there is is a net attractive interaction and so the phase separation
may occur. In the middle region near $j_h/(j_h+U)= 0.5$, a cancellation
of $\tilde{J}_{1}$ and $\tilde{J}_{2}$ leads to a very small net
interaction so that the non-uniform charge phase can not be formed.
Thus, the phase diagram shown in Fig. 2 is determined by two types
of competitions: one competition between hybridization and interaction
and the other between the repulsive and attractive interactions,
respectively, arising from virtual processes (a) and (b).

\begin{figure}
\epsfxsize=8.5cm
\epsfbox{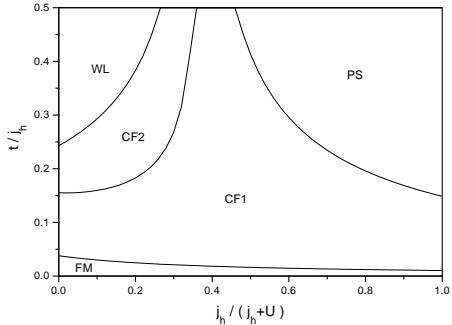}
\caption{
Phase diagram of $t/j_h$ and $j_h/U$. The AF coupling constant
is taken to be $j_{af}/t=0.03$. PS stands for the phase separation. 
}
\end{figure}

At this stage, we wish to point out that in realistic manganites, the
charge ordering is likely to be accompanied by the orbital ordering and
the lattice distortion, \cite{goodenough55} which is not considered in
the present model, for the main purpose of this work is to examine the
effects of the finite $J_{H}$ and $U$. More virtual processes will be
involved if the orbital degeneracy is taken into account, since there
are more mediate states. Meanwhile the lattice distortion will affect
the virtual processes due to its effect on the electron hopping.
These additional virtual processes may be important in the 
realistic manganites. 

Finally, the present method is not confined to the half-doping case. Fig. 3 is the phase diagram for arbitrary doping, in which, for simplicity, $U$
is set to zero to maximize the effect of process (b). In this case, it is
found that the phase separation occurs at either high or low concentrations
of the $e_g$ electrons. The phase diagram obtained here is very similar to
the result of Monte Carlo simulations. \cite{yunoki98} 
Similar results were also obtained analytically by other authors 
\cite{arovas98,nagaev97}.
As $U$ is increased,
process (b) will be suppressed, since the energy paid necessarily for the
double occupancy, $U+J_HS$, will become high. The phase separation
disappears when $U$ is large enough. In doped manganites, the on-site
Coulomb interaction $U$ is usually much stronger than the Hund's coupling
$J_HS$. However, if the orbital degeneracy at each site is further taken
into account, the energy paid for the doubly occupancy might be much lower
than $U+J_HS$,  for two $e_g$ electrons at the same site can occupy
two different orbits and so their spins can be parallel to each other
as well as the core spin. In this sense, the virtual process of the double
occupancy may revive.  It is worth mentioning that the coincidence of
our result at $U=0$ with the numerical simulation indicates the reliability
of our method in this system.

\begin{figure}
\epsfxsize=8.5cm
\epsfbox{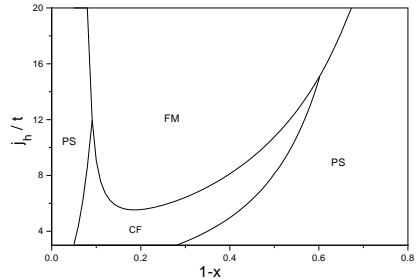}
\caption{
Phase diagram for $j_{af}/t=0.01$ and $U=0$. 
}
\end{figure}

\section{Conclusion}

In summary, we have studied an extended Kondo lattice model in the
presence of strong but finite Hund's coupling and on-site Coulomb
interaction. By means of the Schwinger-boson representation and a
mean-field approximation, we show that the effects of the finite $J_H$ and
$U$ favor antiferromagnetism. In the half-doping case, it is found that the
charge ordering may be superimposed on the magnetic ordering by the repulsive
interaction between electrons due to the virtual process of electron hopping.
If the on-site Coulomb interaction is weak, the phase separation may occur
in either high- or low-doping case. The calculated results show that the
finite $J_H$ and $U$ effects play an important role in forming the magnetic
ordering and non-uniform charge distributions in the doped manganese oxides.

\acknowledgments{ This work was supported by a CRGC 
grant at the University of Hong Kong. One of us (D.Y.X) acknowledges
support from the National Natural Science Foundation of China.

\end{document}